\documentclass[prb,aps,twocolumn,nofootinbib]{revtex4}
\usepackage{graphicx}  \usepackage{dcolumn}  \usepackage{bm} \usepackage{amsmath}  \usepackage{amsfonts}  \usepackage{epsfig} \usepackage{fancyhdr}  \usepackage{enumerate}  \usepackage{float} \usepackage{setspace}  \usepackage{fancybox}

  \newcommand{\Rb}{{\bf R}}  \newcommand{\Ub}{{\bf U}}  \newcommand{\jb}{{\bf j}} 

\begin{document}
\title{Plasmons in electrostatically doped graphene}
\author{Sukosin Thongrattanasiri}
\email{sukosin@gmail.com}
\affiliation{IQFR - CSIC, Serrano 119, 28006 Madrid, Spain}
\author{Iv\'an Silveiro}
\affiliation{IQFR - CSIC, Serrano 119, 28006 Madrid, Spain}
\author{F. Javier Garc\'{\i}a de Abajo}
\email{J.G.deAbajo@csic.es}
\affiliation{IQFR - CSIC, Serrano 119, 28006 Madrid, Spain}

\date{\today}


\begin{abstract}
Graphene has raised high expectations as a low-loss plasmonic material in which the plasmon properties can be controlled via electrostatic doping. Here, we analyze realistic configurations, which produce inhomogeneous doping, in contrast to what has been so far assumed in the study of plasmons in nanostructured graphene. Specifically, we investigate backgated ribbons, co-planar ribbon pairs placed at opposite potentials, and individual ribbons subject to a uniform electric field. Plasmons in backgated ribbons and ribbon pairs are similar to those of uniformly doped ribbons, provided the Fermi energy is appropriately scaled to compensate for finite-size effects such as the divergence of the carrier density at the edges. In contrast, the plasmons of a ribbon exposed to a uniform field exhibit distinct dispersion and spatial profiles that considerably differ from uniformly doped ribbons. Our results provide a road map to understand graphene plasmons under realistic electrostatic doping conditions.
\end{abstract}
\maketitle


Plasmons\cite{R1988} --the collective oscillations of conduction electrons in metals-- are capable of confining electromagnetic energy down to deep sub-wavelength regions. They can also enhance the intensity of an incident light wave by several orders of magnitude. These phenomena are the main reason why the field of plasmonics is finding a wide range of applications that include single-molecule sensing,\cite{paper125} nonlinear optics,\cite{DN07} and optical trapping of nanometer-sized objects.\cite{JRQ11}

Recently, confined plasmons have been observed and spatially mapped in doped graphene.\cite{graphene1st} The level of doping in this material can be adjusted by exposing it to the electric fields produced by neighboring gates. Electrostatic doping has actually been used to demonstrate plasmon-frequency tunability\cite{graphene1st} and induced optical modulations in the THz\cite{JGH11} and infrared\cite{FAB11} response of graphene.

The two-dimensional (2D) band structure of pristine graphene consists of two cones filled with valence electrons and two empty inverted cones joining the former at the so-called Dirac points, which mark the Fermi level. Extra electrons or holes added to this structure form a 2D electron or hole gas that can sustain surface plasmons.\cite{S1986,WSS06} Compared to noble-metal plasmons, graphene modes are believed to be long-lived excitations.\cite{JBS09} But most importantly, their frequency can be controlled via the above-mentioned electrostatic doping.\cite{JGH11,FAB11,graphene1st} For example, in homogeneous suspended graphene, a perpendicular DC electric field $\mathcal{E}$ applied to one side of the carbon sheet is completely screened by an induced surface charge density $-en=\mathcal{E}/4\pi$, and this in turn situates the Fermi level at an energy $E_F=\hbar v_F \sqrt{\pi|n|}$ relative to the Dirac points.\cite{CGP09} Here, $v_F\approx10^6\,$m/s is the Fermi velocity of graphene.

\begin{figure}
\begin{center}
\includegraphics[width=85mm,angle=0,clip]{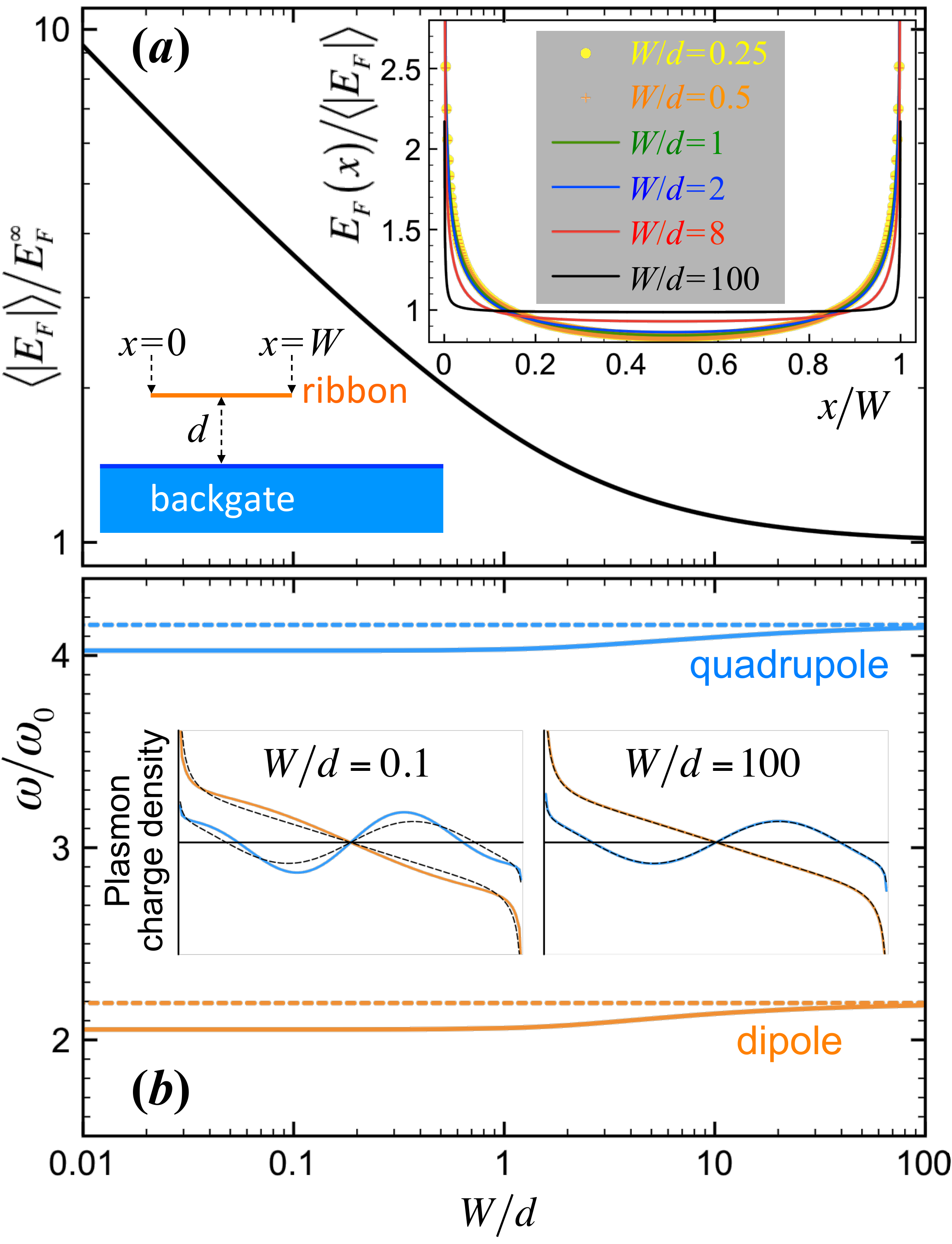}
\caption{Electrostatic doping and plasmon modes in backgated graphene ribbons. {\bf (a)} Average Fermi energy $\langle|E_F|\rangle$ as a function of width-to-distance ratio $W/d$, normalized to the value $E_F^\infty$ obtained in the $W\gg d$ limit. The upper inset shows the $E_F$ distribution across the ribbon, normalized to $\langle|E_F|\rangle$. The lower inset shows a sketch of the geometry. {\bf (b)} Frequency $\omega$ of the dipolar and quadrupolar plasmons, normalized to $\omega_0=(e/\hbar)\sqrt{\langle|E_F|\rangle/W}$, as obtained from the Drude model. The insets show the surface charge-density oscillating at frequency $\omega$ and corresponding to these plasmons (vertical axis) as a function of position across the ribbon (horizontal axis). The dashed curves indicate the $W\gg d$ limit.} \label{Fig1}
\end{center}
\end{figure}

Plasmons in doped graphene nanostructures have been generally studied by assuming a uniform doping electron density $n$.\cite{VE11,paper176} But in practice $n$ is inhomogeneous and depends on the actual geometrical configuration. For example, in a ribbon of width $W$ placed at a distance $d$ from a planar biased backgate, $n$ shows a dramatic pileup near the edges, as it is well known in microstrip technology,\cite{W1964,W1977,SE08,VZ10} leading to divergent local $E_F$ levels, as shown in the inset of Fig.\ \ref{Fig1}(a) for various values of $W/d$. More precisely, $n$ diverges as $\propto1/\sqrt{x}$ with the distance to the edge $x$ (and hence, $E_F\propto1/x^{1/4}$). The average of the Fermi energy over the ribbon area [$\langle|E_F|\rangle$, see Fig.\ \ref{Fig1}(a)] is very different from the $W/d\gg1$ limit ($E_F^\infty=\hbar v_F\sqrt{|V|/4ed}$) and diverges as\cite{analimit} $\sqrt{d/W}/\log(2d/W)$ in the narrow ribbon limit ($W\ll d$) for constant bias potential $V$. Therefore, the question arises, how different are the plasmon energies and field distributions in actual doped graphene nanostructures compared to those obtained for uniformly doped graphene?

Here, we analyze plasmons in doped graphene ribbons under different geometrical configurations. Specifically, we study backgated single ribbons, co-planar parallel ribbon pairs of opposite polarity, and single ribbons immersed in a uniform external electric field. For simplicity, we describe the frequency-dependent conductivity of doped graphene in the Drude model as $\sigma(\omega)=(ie^2|E_F|/\pi\hbar^2)/(\omega+i\gamma)$, where $\gamma\ll\omega$ is a small relaxation rate. The doping electron density $n$ is obtained from electrostatic boundary-element calculations, while the plasmon frequencies are computed using a discrete surface-dipole approximation (DSDA), as explained in Appendix\ \ref{appcomputation}.

{\it Backgated ribbons.-} A first conclusion extracted from Fig.\ \ref{Fig1}(a) is that the level of doping, quantified in the Fermi energy, is not well described by the simple capacitor analysis of the $W\gg d$ geometry. Normalizing to the average value $\langle|E_F|\rangle$, we find Fermi-energy profiles that vary between a well of sharp corners for large $W/d$ and a smoother, converged shape for small $W/d$ [upper inset of Fig.\ 1(a)]. The former limit corresponds to the ribbon in close proximity to the backgate, in which $E_F$ is nearly uniform. In contrast, at large separations ($d\gg W$) we find a profile determined by the interaction with a distant image, which converges to a well-defined shape up to an overall factor $\langle |E_F|\rangle$ evolving as shown in the main plot of Fig.\ 1(a).

It is convenient to normalize the ribbon plasmon frequencies to $\omega_0=(e/\hbar)\sqrt{\langle|E_F|\rangle/W}$, so that $\omega/\omega_0$ is a dimensionless number, independent of the specific ribbon width $W$ and gate voltage (i.e., $\langle|E_F|\rangle$), as proved in Appendix\ \ref{appscaling}. For example, with $W=100\,$nm and $\langle|E_F|\rangle=0.5\,$eV, we find $\hbar\omega_0=0.085\,$eV and a dipole plasmon energy $\hbar\omega\sim0.17\,$eV (wavelength $\sim7.3\,\mu$m). With this normalization, $\omega/\omega_0$ shows just a mild dependence on $W/d$ for the dipolar and quadrupolar modes [Fig.\ \ref{Fig1}(b)]. The corresponding induced densities (insets) are only slightly affected by the change in doping profile relative to uniform doping (i.e., the average level of doping is a dominant parameter, and the effect of edge divergences is only marginal). In conclusion, the plasmon frequencies and induced densities can be approximately described by assuming a uniform Fermi energy in backgated ribbons, thus supporting the validity of previous analyses for this configuration,\cite{paper176,NGG11,paper181} although the Fermi energy has to be appropriately scaled as shown in Fig.\ \ref{Fig1}(a) to compensate for the effect of finite $W/d$ ratios.

\begin{figure}
\begin{center}
\includegraphics[width=85mm,angle=0,clip]{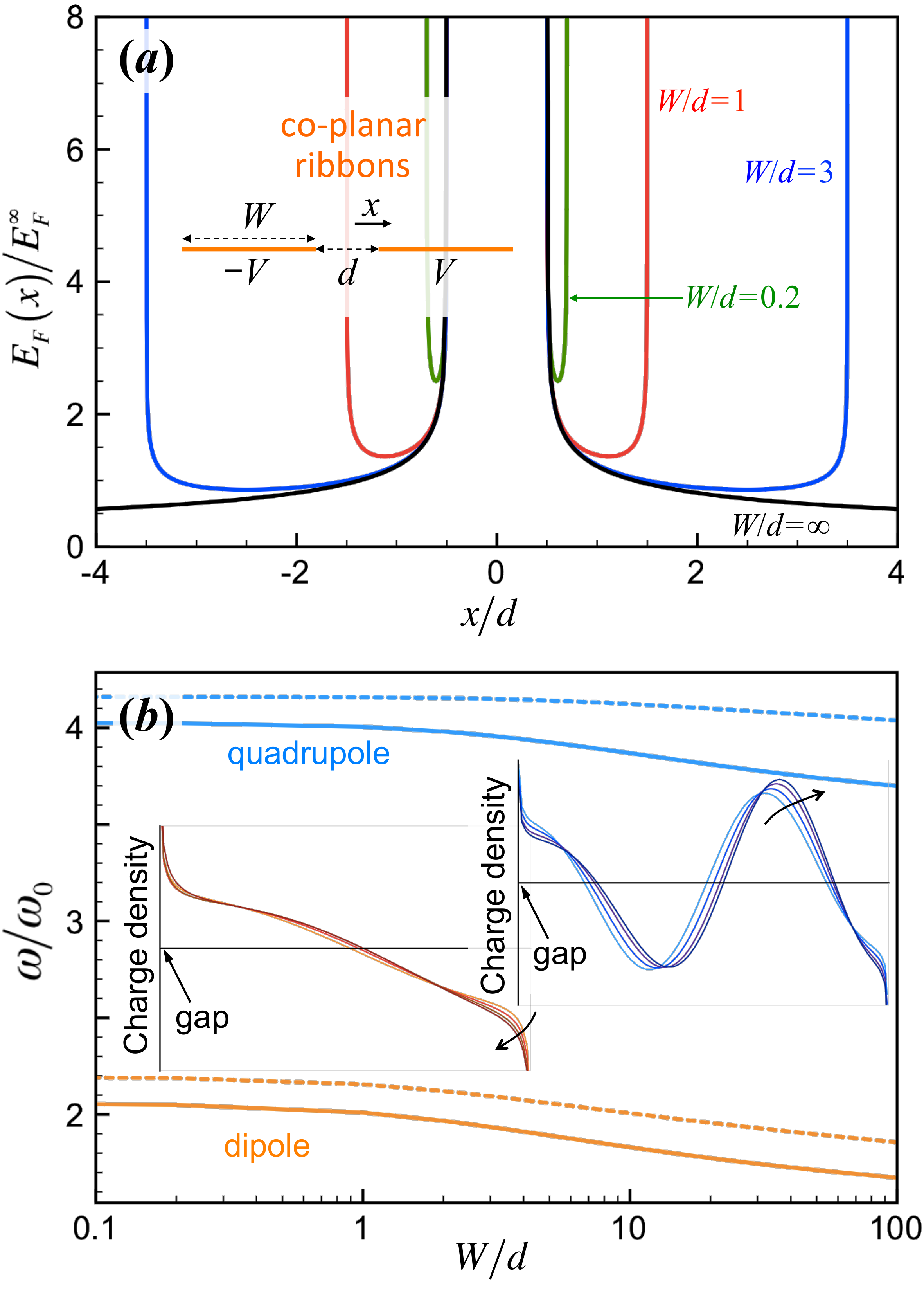}
\caption{Plasmons in pairs of co-planar parallel graphene ribbons of opposite polarity. {\bf (a)} Fermi energy distribution across pairs of ribbons for different ratios of the ribbon width $W$ to the gap distance $d$. The ribbons are placed at potentials $-V$ and $V$, respectively (see inset), and the Fermi energy $E_F$ is normalized to the value $E_F^\infty=\hbar v_F\sqrt{|V|/4ed}$. {\bf (b)} Frequency $\omega$ of the dipolar and quadrupolar modes as a function of $W/d$, normalized to $\omega_0=(e/\hbar)\sqrt{\langle|E_F|\rangle/W}$. Solid (dashed) curves correspond to inhomogeneous (uniform) doping. The insets show the plasmon charge-density associated with both modes (vertical axis) as a function of position across the ribbon on the right (horizontal axis, with the position of the gap indicated by an arrow) for $W/d=0.2,1,3,$ and 10 (curves evolving in the direction of the arrows).} \label{Fig2}
\end{center}
\end{figure}

{\it Two co-planar parallel graphene ribbons.-} Two neighboring ribbons can act both as plasmonic structures and as gates. We explore this possibility in Fig.\ \ref{Fig2}, where the ribbons are taken to be oppositely polarized. This produces doping profiles as shown in Fig.\ \ref{Fig2}(a), which evolve from a shape similar to the one obtained for the single ribbon of Fig.\ \ref{Fig1} in the small-ribbon limit at large ribbon-pair separations, towards a converged profile near the gap in the $W/d\rightarrow\infty$ limit. Again, plasmons in this structure are very similar to those of neighboring uniformly doped ribbons [see Fig.\ \ref{Fig2}(b)], provided one compares for the same value of the average Fermi energy $\langle|E_F|\rangle$. The separation dependence of $\langle|E_F|\rangle$ is shown in Appendix\ \ref{calculationof}. Incidentally, plasmons in pairs of uniform ribbons have been thoroughly described and the evolution of the plasmon frequency with distance explained in a recent publication,\cite{paper181} including the redshift with decreasing $d$.

\begin{figure}
\begin{center}
\includegraphics[width=85mm,angle=0,clip]{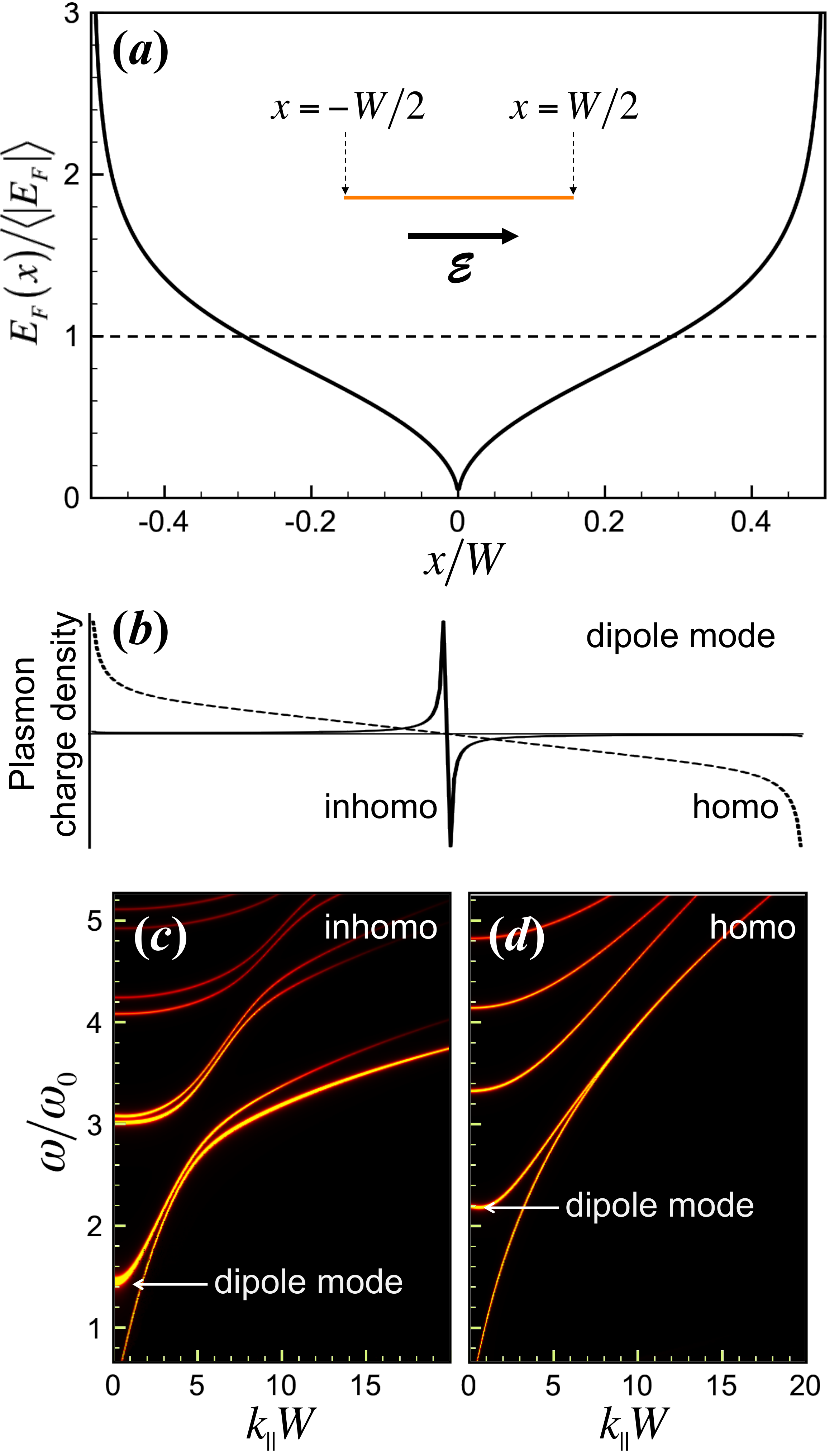}
\caption{Plasmons in individual ribbons subject to a uniform external electric field $\mathcal{E}$. {\bf (a)} Fermi energy distribution normalized to $\langle|E_F|\rangle=0.6\,\hbar v_F\sqrt{|\mathcal{E}|/e}$. The inset shows a sketch of the geometry. {\bf (b)} Surface charge associated with the dipolar plasmon mode (solid curve). The plasmon frequency is $\omega\approx1.45\,\omega_0$, where $\omega_0=(e/\hbar)\sqrt{\langle|E_F|\rangle/W}$. The dashed curve shows the charge density profile for a uniform doping density (i.e., $E_F=\langle|E_F|\rangle$). {\bf (c)} Plasmon dispersion diagram representing the dependence of the density of optical states on frequency $\omega$ and wave vector parallel to the ribbon $k_\parallel$. {\bf (d)} Same as (c) for uniform doping.} \label{Fig3}
\end{center}
\end{figure}

{\it Doping through a uniform electric field.-} A disadvantage of the above doping schemes is the fabrication process involved in adding contacts that allow electrically charging the graphene, which can be a source of  defects in the carbon layer. This could be avoided by doping through an external electric field $\mathcal{E}$ produced by either distant gates or low-frequency radiation. The graphene would then remain globally neutral. This possibility is analyzed in Fig.\ \ref{Fig3}, where we consider a ribbon subject to a uniform electric field directed across its width. The doping profile [Fig.\ \ref{Fig3}(a)] is again exhibiting a divergence of $E_F$ at the edges, and it vanishes at the center of the ribbon, where the doping density changes sign (see Appendix\ \ref{ribbonin}). Following the methods described in Appendix\ \ref{ribbonin}, we find the relation $\langle|E_F|\rangle=0.6\,\hbar v_F\sqrt{|\mathcal{E}|/e}$. The resulting dipolar plasmon [Fig.\ \ref{Fig3}(b), solid curve] displays a large concentration of induced charges near the center of the ribbon, in contrast to the dipole plasmon obtained for a homogeneous doping density (dashed curve). This inhomogeneous dipole-charge concentration is induced by the vanishing of the doping charge density, which can be understood as a thinning of the effective graphene-layer thickness, similar to what happens near the junction of two barely touching metallic structures (e.g., two spheres\cite{paper075}).

We have so far discussed plasmons that are invariant along the length of the ribbon (i.e., as those excited by illuminating with light incident perpendicularly to the graphene). But now, we show in Fig.\ \ref{Fig3}(c),(d) the full plasmon dispersion relation as a function of frequency $\omega$ and wave vector $k_\parallel$ parallel to the ribbon, both for inhomogeneous doping produced by an external uniform field [Fig.\ \ref{Fig3}(c)] and for a ribbon with uniform doping density [Fig.\ \ref{Fig3}(d)]. The dispersion relations are rather different in both situations, with the inhomogeneous ribbon showing a denser set of modes, as well as more localization in the lowest-energy plasmons for large $k_\parallel$, as we show in Appendix\ \ref{appcomputation} by means of near-field plots for the lowest-energy modes of both types of ribbons.

Finally, let us mention that the inelastic plasmon decay rate is given by $\gamma$ within the Drude model in uniformly doped structures.\cite{paper181} However, $\gamma$ depends on position for inhomogeneous doping. Using the DC mobility $\mu$, one can estimate $\gamma=ev_F^2/\mu|E_F(x)|\approx2\times10^{12}\,$s$^{-1}$ for $E_F=0.5\,$eV and a typical measured mobility $\mu=10,000\,$cm$^2/$Vs.\cite{NGM04,ZTS05} Noticing that the local contribution to inelastic losses is proportional to ${\rm Re}\{\sigma\}\approx(e^2/\pi\hbar^2)|E_F|\gamma/\omega^2$ (i.e., independent of $x$), we conclude that the inhomogeneity of $\gamma$ is however translated into a uniform spatial distribution of losses.

{\it Conclusions.-} We have shown that the plasmons of doped graphene ribbons are highly sensitive to the inhomogeneities of the doping charge density produced by realistic electrostatic landscapes. The doping profile can be engineered by adjusting the configuration of the gates relative to the graphene. We find an interesting scenario when a uniform external electric field is used to dope the graphene, leading to plasmons with very different characteristics (e.g., induced charges piling up near the center of the ribbon) compared to those of uniformly doped graphene (in which plasmons pile up at the edges). This configuration can be used to avert losses associated with nonlocal effects at the edges, which are expected to be significant.\cite{paper183} The present study can be straightforwardly extended to other configurations, such as finite graphene nanoislands exposed to either backgates or side gates. Electrostatic charge accumulation at sharp edges can offer an additional handle to manipulate plasmon modes. In addition to the possibilities explored in this paper, one can use biased tips to produce localized disk-like doping areas at designated positions targeted by simply moving the tips above a graphene flake. In conclusion, the design of electrostatic landscapes becomes a useful tool to engineer graphene plasmons. 

We would like to thank Enrique Bronchalo for helpful discussions. This work has been supported by the Spanish MICINN (MAT2010-14885 and Consolider NanoLight.es) and the European Commission (FP7-ICT-2009-4-248909-LIMA and FP7-ICT-2009-4-248855-N4E).

\appendix

\begin{figure*}
\begin{center}
\includegraphics[width=170mm,angle=0,clip]{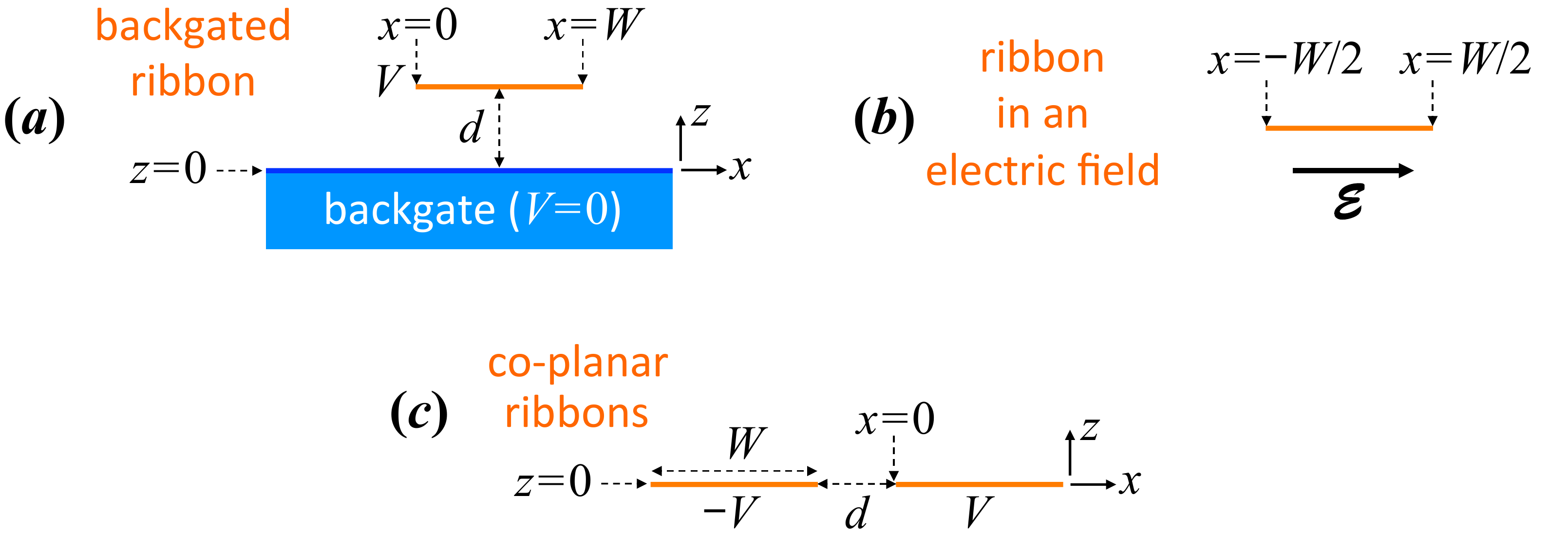}
\caption{Geometries under consideration for electrostatic doping of graphene: (a) single ribbon placed at a potential $V$ relative to a backgate; (b) single ribbon exposed to a uniform external electric field; (c) two co-planar parallel ribbons set at opposite potentials, $-V$ and $V$.} \label{FigSM1}
\end{center}
\end{figure*}

\begin{figure}
\begin{center}
\includegraphics[width=85mm,angle=0,clip]{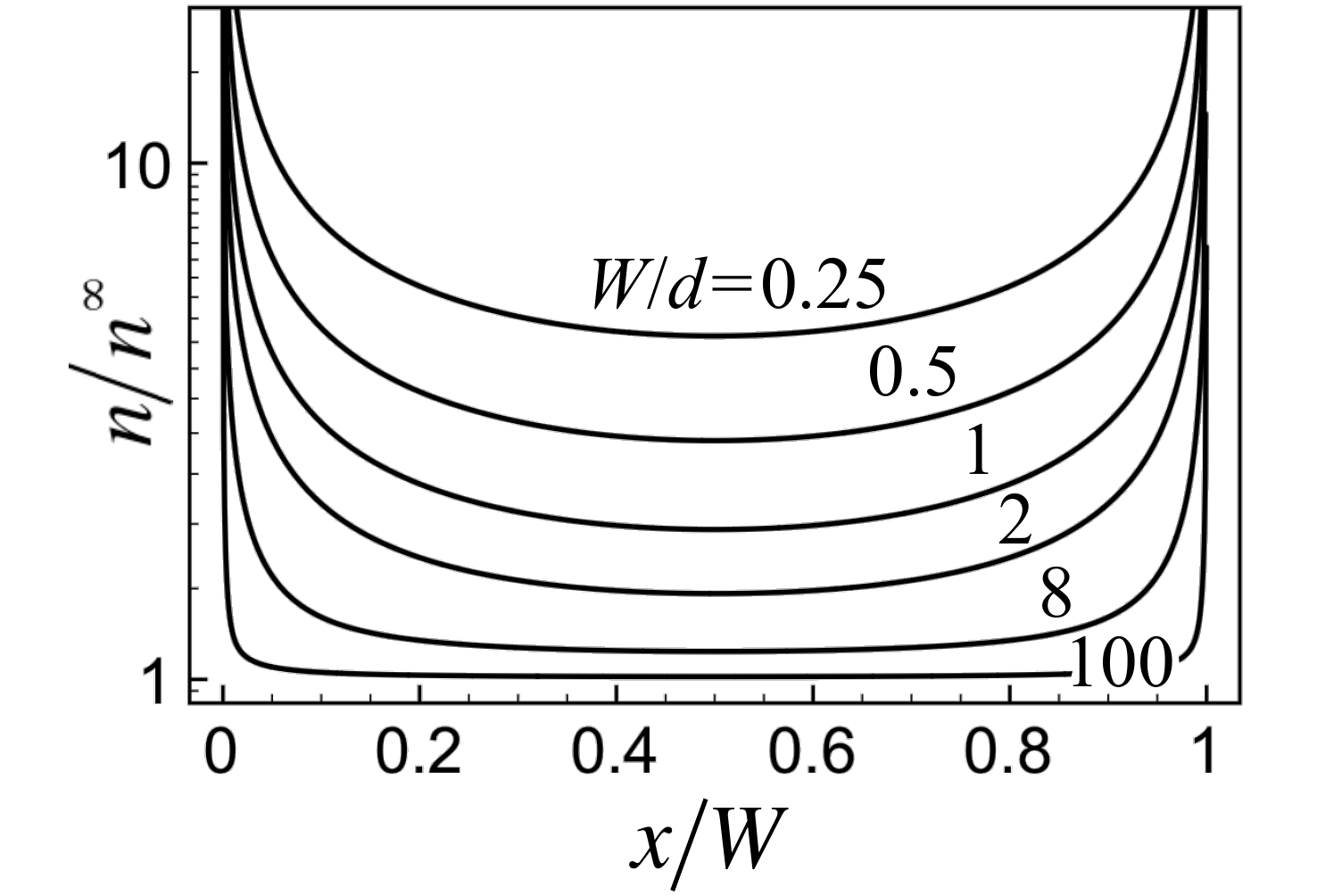}
\caption{Electrostatic doping electron density in backgated ribbons under the configuration sketched in Fig.\ \ref{FigSM1}(a) for different values of the ratio of the ribbon width $W$ to the gap distance $d$. The electron density $n$ is normalized to the value $n^\infty=-V/4\pi ed$ corresponding to the $W/d\gg1$ limit.} \label{FigSM4}
\end{center}
\end{figure}

\begin{figure}
\begin{center}
\includegraphics[width=85mm,angle=0,clip]{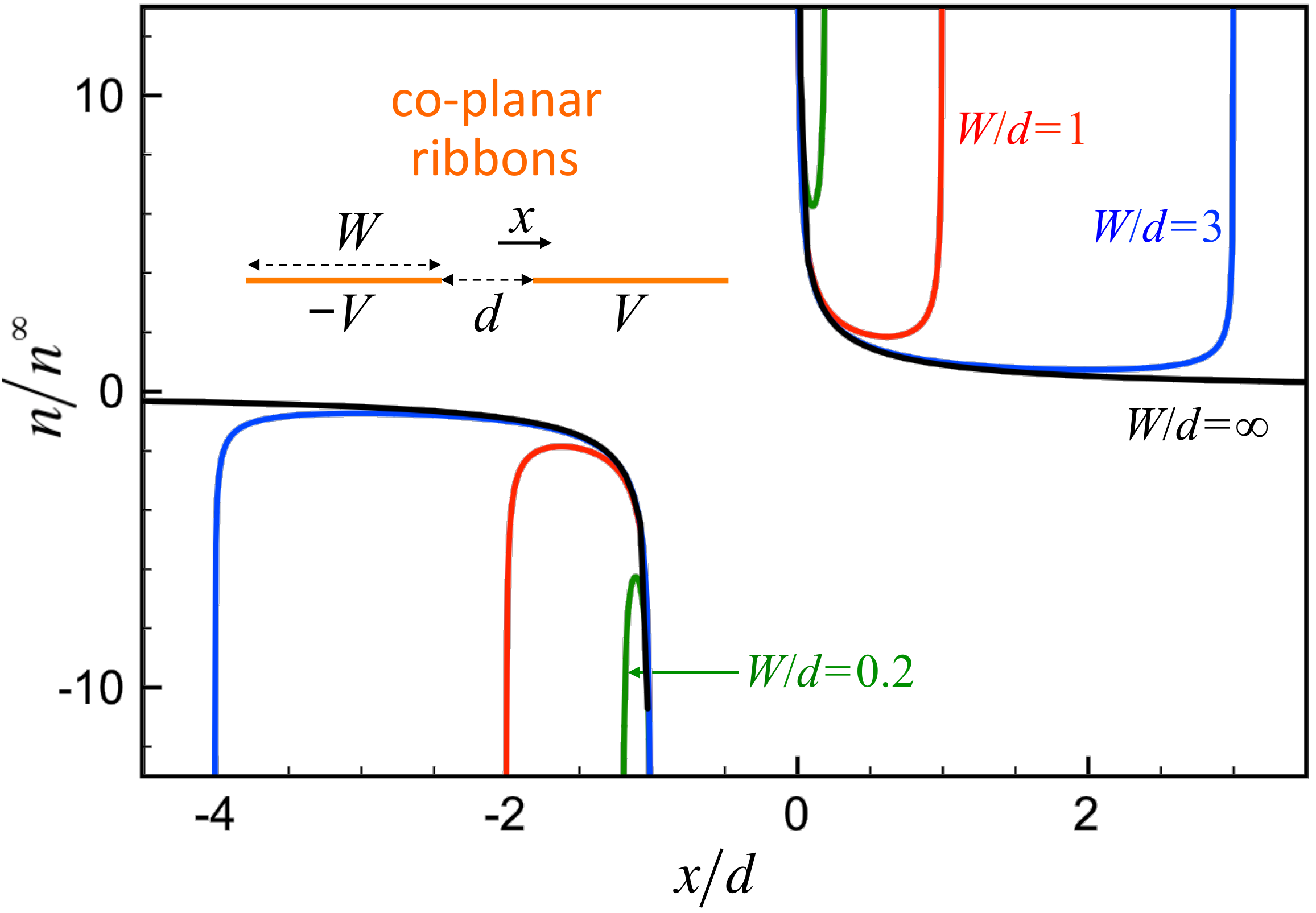}
\caption{Electrostatic doping electron density across pairs of ribbons for different ratios of the ribbon width $W$ to the gap distance $d$. The ribbons are placed at potentials $-V$ and $V$, respectively (see inset), and the electron density $n$ is normalized to the value $n^\infty=-V/4\pi ed$.} \label{FigSM3}
\end{center}
\end{figure}

\begin{figure}
\begin{center}
\includegraphics[width=85mm,angle=0,clip]{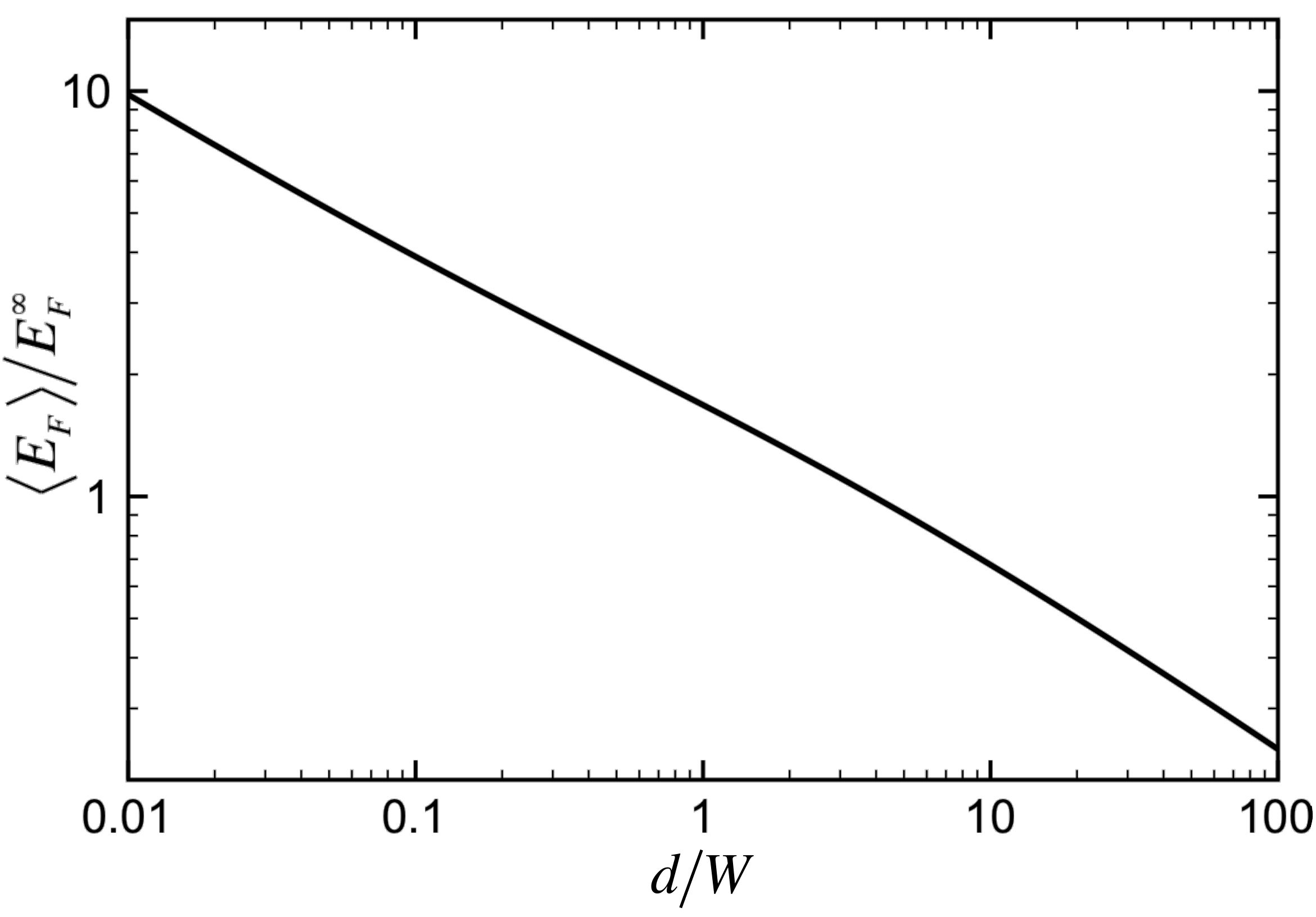}
\caption{Average Fermi energy $\langle |E_F|\rangle$ as a function of $d/W$ for pairs of ribbons under the configuration of Fig.\ \ref{FigSM1}(c). The Fermi energy is normalized to $E_F^\infty=\hbar v_F\sqrt{|V|/4ed}$.} \label{FigSM5}
\end{center}
\end{figure}

\section{Electrostatic calculation of doping surface charge distributions}
\label{calculationof}

We consider three different geometries for electrically doping graphene, as illustrated in Fig.\ \ref{FigSM1}. Geometries in Figs.\ \ref{FigSM1}(a),(c) refer to doping of biased graphene ribbons, which accumulate a net electric charge. In contrast, the ribbon of Fig.\ \ref{FigSM1}(b), exposed to an external uniform field, has zero total charge. The charge distribution in a backgated ribbon [Fig.\ \ref{FigSM1}(a)] has been reported in the past,\cite{W1964,W1977,SE08,VZ10} particularly in the context of microstrip technology,\cite{W1964,W1977} but we provide here a general procedure to calculate it based upon boundary elements, which we extend to other geometries under consideration.

\begin{widetext}

\subsection{Backgated graphene ribbon}
\label{backgatedribbon}

A ribbon placed at a potential $V$ relative to a backgate [Fig.\ \ref{FigSM1}(a)] displays a charge density $-en(x)$ that can be calculated using the method of images (notice that $n$ is the doping electron density). The charge depends on the coordinate across the ribbon $x$, which varies in the range $0<x<W$, where $W$ is the width. Using the method of images, this problem is equivalent to two parallel ribbons vertically separated by a distance $2d$ and placed at potentials $V$ (upper ribbon) and $-V$ (lower ribbon), so that the backgate plane ($z=0$) is at zero potential. The lower ribbon is thus represented by a charge density $en(x)$. The potential at $x$ in the upper ribbon is then given by
\begin{equation}
V=\int_0^W dx'\int_{-\infty}^\infty dy \,[-en(x')]\, \left[\frac{1}{\sqrt{(x-x')^2+y^2}}-\frac{1}{\sqrt{(x-x')^2+y^2+4d^2}}\right].\label{self1}
\end{equation}
\end{widetext}
Analytically performing the integral along the $y$ coordinate (perpendicular to the plane of Fig.\ \ref{FigSM1}) and using the notation $\eta=W/d$, $\theta=x/d$, and \[u=-V/ed,\] Eq.\ (\ref{self1}) reduces to
\begin{equation}
u=\int_0^\eta d\theta' \,n(\theta'd)\, F(\theta,\theta'), \label{self}
\end{equation}
where
\begin{equation}
F(\theta,\theta')=\ln\left[1+\frac{4}{(\theta-\theta')^2}\right].\nonumber
\end{equation}
We solve this integral equation by discretizing $\theta$ through a set of $N$ equally spaced points $\theta_j=(j+1/2)\eta/N$, with $j=0,\dots,N-1$. Equation\ (\ref{self}) is then approximated as
\begin{equation}
u\approx\sum_{j'} \,n(\theta_{j'}d)\, M_{jj'}, \label{uu}
\end{equation}
where
\begin{equation}
M_{jj'}=\int_{\theta_{j'}-\eta/2N}^{\theta_{j'}+\eta/2N} d\theta' \, F(\theta_j,\theta') \label{MM}
\end{equation}
is an integral over the interval surrounding point $\theta_{j'}$. We have assumed that $n(x)$ is a smooth function, and although we find later that it diverges as $\propto1/\sqrt{x}$ near the edges (see main paper), this divergence is integrable and contributes negligibly to the total integral for $N\gg1$. From here, the charge distribution is found by inverting the matrix $M$, so that
\begin{equation}
n(\theta_jd)=u\sum_{j'} \left[M^{-1}\right]_{jj'}. \label{solution}
\end{equation}
In practice, this method converges for $N\sim100$. Similar convergence is obtained for the geometries considered in Secs.\ \ref{twoco} and \ref{ribbonin}. Each curve in Fig.\ 1(a) of the main paper is actually consisting of two curves obtained with $N=100$ and $N=500$, and one cannot tell the difference between them on the scale of the plot. Finally, notice that the uniform electron density in the $W/d\gg1$ limit is given by $n^\infty=u/4\pi$. Also, we obtain the average of the Fermi energy $\langle |E_F|\rangle$ normalized to the $W/d\gg1$ limit $E_F^\infty=\hbar v_F\sqrt{|u|/4}$ as
\begin{equation}
\frac{\langle |E_F|\rangle}{E_F^\infty}\approx\frac{1}{N}\sum_j\sqrt{4\pi |n(\theta_jd)/u|}, \nonumber
\end{equation}
where $v_F\approx10^6\,$m/s is the Fermi velocity. (Incidentally, we consider the absolute value of $E_F$ because the graphene response is nearly insensitive to the sign of $E_F$.)

Figure\ \ref{FigSM4} shows examples of doping charge densities from which we have extracted the Fermi energies shown in the inset of Fig.\ 1(a) of the main paper.

\subsection{Two co-planar parallel ribbons set at opposite potentials}
\label{twoco}

We can repeat the same analysis as in Sec.\ \ref{backgatedribbon} for two ribbons arranged as shown in Fig.\ \ref{FigSM1}(c) and set at potentials $-V$ (left ribbon) and $V$ (right ribbon). The separation between ribbons is $d$, and $x=0$ is chosen at the left edge of the right ribbon. Equations\ (\ref{self})-(\ref{solution}) remain valid when we consider the charge distribution in the right ribbon, but now the kernel of Eq.\ (\ref{self}) becomes
\begin{equation}
F(\theta,\theta')=2\ln\left|\frac{\theta+\theta'+1}{\theta-\theta'}\right|.\nonumber
\end{equation}
The charge density in the left ribbon is found from the symmetry $n(x)=-n(-x-d)$.

Figure\ \ref{FigSM3} shows examples of doping charge densities from which we have extracted the Fermi energies shown in Fig.\ 2(a) of the main paper. The average Fermi energy $\langle |E_F|\rangle$ is shown in Fig.\ \ref{FigSM5} as a function of $d/W$.

\begin{figure}
\begin{center}
\includegraphics[width=85mm,angle=0,clip]{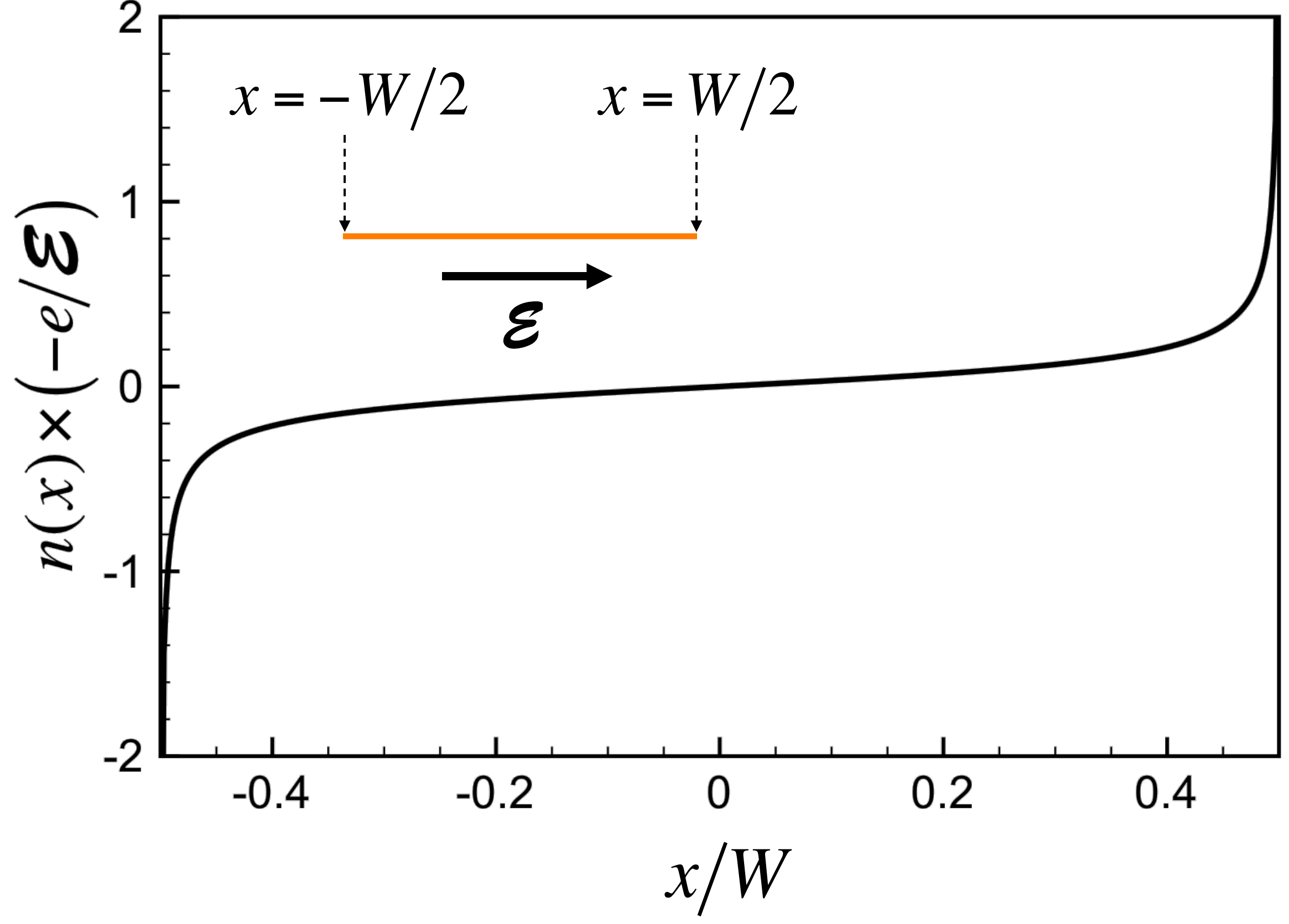}
\caption{Doping charge density in a graphene ribbon subject to an external uniform electric field $\mathcal{E}$.} \label{FigSM2}
\end{center}
\end{figure}

\subsection{Graphene ribbon in a uniform electric field}
\label{ribbonin}

For a ribbon subject to an external uniform electric field, only the component parallel to the ribbon $\mathcal{E}$ can produce charge redistributions and local doping [see Fig.\ \ref{FigSM1}(b)]. Following the same procedure as in Sec.\ \ref{backgatedribbon}, and taking the ribbon to be placed at zero potential, we can write
\begin{equation}
0=-\mathcal{E}x+\int_{-W/2}^{W/2} dx'\int_{-\infty}^\infty dy \,[-en(x')]\, \frac{1}{\sqrt{(x-x')^2+y^2}},\nonumber
\end{equation}
where the first term is the scalar potential produced by the external field. Now, using the normalization $\theta=x/W$, the above equation reduces to
\begin{equation}
\mathcal{E}\theta=\int_{-1/2}^{1/2} d\theta' \,[-en(\theta'W)]\, F(\theta,\theta'), \nonumber
\end{equation}
where
\begin{equation}
F(\theta,\theta')=\ln\left[\frac{1}{(\theta-\theta')^2}\right].\nonumber
\end{equation}
We use the discretization $\theta_j=-1/2+(j+1/2)/N$ to write an expression similar to Eq.\ (\ref{uu}), where $M_{jj'}$ is still given by Eq.\ (\ref{MM}) with $\eta=1$. Finally, the electron density is obtained from
\begin{equation}
n(\theta_jd)=(-\mathcal{E}/e)\sum_{j'} \left[M^{-1}\right]_{jj'}\theta_{j'}. \nonumber
\end{equation}
The doping charge density obtained from this equation is represented in Fig.\ \ref{FigSM2}. From here, we find a local Fermi energy $E_F(x)=\hbar v_F\sqrt{\pi|n(x)|}$ as shown in Fig.\ 3(a) of the main paper. The average Fermi energy is found to be $\langle |E_F|\rangle=0.6\,\hbar v_F\sqrt{|\mathcal{E}|/e}$.



\section{Scaling of plasmon frequencies in the Drude model}
\label{appscaling}

The electric scalar potential $\phi$ associated with the plasmon modes in a doped graphene planar nanostructure satisfies the self-consistent electrostatic equation
\begin{equation}
\phi(\Rb)=\frac{i\chi}{\omega}\int \frac{d^2\Rb'}{|\Rb-\Rb'|}\,\nabla\cdot\left[\sigma(\Rb',\omega)\,\nabla\phi(\Rb')\right],\label{self2}
\end{equation}
where an $\exp(-i\omega t)$ time dependence is undertood and all quantities are evaluated at the graphene plane (i.e., coordinates $\Rb$ and $\Rb'$ are in that plane). This equation represents the field induced by the self-consistent surface-charge density, which is calculated from the continuity equation as $(-i/\omega)\nabla\cdot\jb$ in terms of the current $\jb$, and this is in turn proportional to the in-plane electric field via $-\sigma\nabla\phi$, where $\sigma$ is the conductivity. Incidentally, $\sigma=0$ outside the graphene, so that $\sigma$ presents a jump at the graphene edge, the gradient of which generates an edge charge density. Equation\ (\ref{self2}) is valid for small patterning lengths compared to the light wavelength. The constant $\chi$ takes the values $\chi=1$ for suspended graphene in vacuum, $\chi=1/\epsilon$ for graphene embedded in a uniform dielectric of permittivity $\epsilon$, or $\chi=2/(\epsilon+1)$ for graphene supported on a substrate.\cite{paper181} Now, we consider the Drude model for the conductivity,
\begin{equation}
\sigma(\Rb,\omega)=\frac{e^2}{\pi\hbar^2}\frac{i|E_F(\Rb)|}{(\omega+i\gamma)},\label{Drude}
\end{equation}
where $0<\gamma\ll\omega$ and the $\Rb$ dependence comes from the inhomogeneous electron density $n(\Rb)$ (see Sec.\ \ref{calculationof}), which locally situates the Dirac points at an energy $E_F(\Rb)=\hbar v_F\sqrt{\pi|n(\Rb)|}$ relative to the Fermi level.\cite{CGP09}

From the analysis of Sec.\ \ref{calculationof}, we can write the Fermi energy distribution as $E_F(\Rb)=\langle |E_F|\rangle\,f(\Rb)$, where $f$ is a dimensionless envelope function. Likewise, distances can be scaled with a characteristic length $W$ as $\Ub=\Rb/W$. Then, Eq.\ (\ref{self2}) becomes
\begin{equation}
\phi(\Ub)=\zeta\int \frac{d^2\Ub'}{|\Ub-\Ub'|}\,\nabla_{\Ub'}\cdot\left[f(\Ub')\,\nabla_{\Ub'}\phi(\Ub')\right],\nonumber
\end{equation}
where
\begin{equation}
\zeta=\frac{-e^2\chi}{\pi\hbar^2W}\frac{\langle |E_F|\rangle}{\omega(\omega+i\gamma)}\nonumber
\end{equation}
is a dimensionless eigenvalue. The modes of the system satisfy this equation for specific choices of $\zeta$, and therefore, we conclude that the plasmon energies $\omega$ can be naturally normalized to the frequency
\begin{equation}
\omega_0=(e/\hbar)\sqrt{\chi \langle |E_F|\rangle/W}. \label{w0}
\end{equation}
Obviously, $\omega/\omega_0$ does not depend on the specific choice of $\langle |E_F|\rangle$ (the doping level), $W$ (the size of the system), or $\chi$ (the dielectric environment), and consequently, we present results normalized in this way in the main paper, which are universal for the kind of geometries under consideration, provided we stay within the limits of validity of the Drude model (i.e., $\omega<E_F$, and $E_F$ smaller than the optical phonon energy $\sim0.2\,$eV\cite{JBS09}). Additionally, because the electrostatic eigensystem is Hermitian,\cite{OI1989} the eigenvalues $\zeta$ are real, and therefore, the plasmon frequencies have imaginary part ${\rm Im}\{\omega\}=-\gamma/2$, so that the plasmon lifetime is $1/\gamma$, independent of geometrical and physical parameters within the Drude approximation.


\section{Computation of plasmon frequencies and near fields}
\label{appcomputation}

We solve the electrostatic problem of Eq.\ (\ref{self2}) by describing the graphene as a periodic array of surface dipoles with a small period compared to the characteristic lengths of the structure. This is the discrete surface-dipole approximation\cite{DSDA} (DSDA), in which the polarizability of each element is taken such that a layer formed by a uniform lattice of dipoles has the same conductivity as a uniform layer of graphene. The sum over dipole elements along $y$ is performed before a self-consistent solution is sought, and therefore, the numerical problem reduces to solving a set of $2N$ linear equations with $2N$ variables (the dipole components along both $x$ and $y$ directions), where $N$ is the number of dipoles across $x$. In practice, convergence is achieved with a few hundred dipoles for the dimensions considered in this work. Here, we have modified this method by allowing each element to depend through $E_F$ on the spatial position along $x$.

\begin{figure*}
\begin{center}
\includegraphics[width=110mm,angle=0,clip]{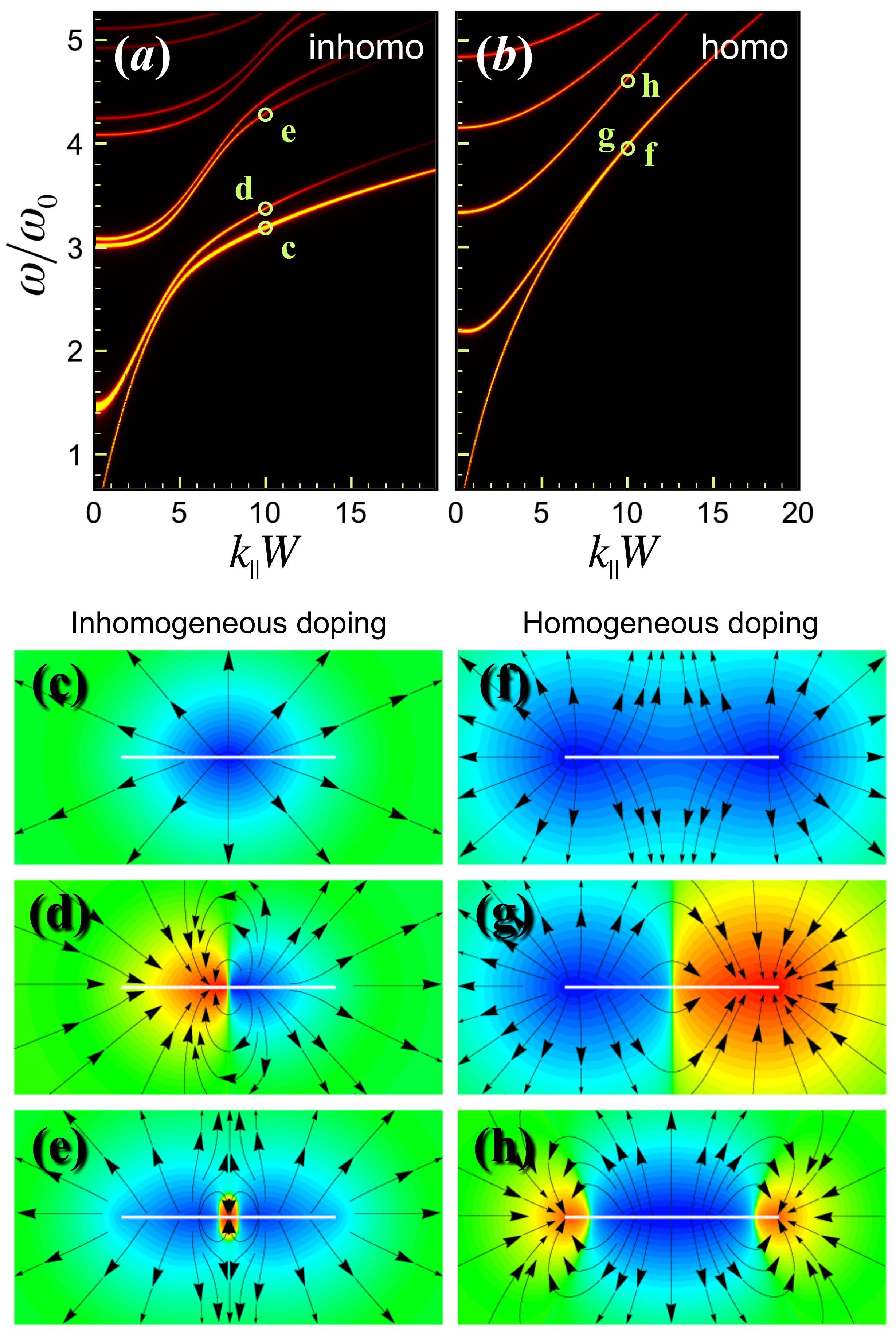}
\caption{{\bf (a)} Plasmon dispersion diagram representing the dependence of the density of optical states on frequency $\omega$ and wave vector parallel to the ribbon $k_\parallel$ for doping produced by a uniform external electric field (see Fig.\ \ref{FigSM2}). {\bf (b)} Same as (a) for uniform doping. {\bf (c)-(h)} Near-field plots showing the $y$ electric field component along the direction of the ribbon translational symmetry for the modes signaled by open circles in (a) and (b). The plane of the figure is perpendicular to the ribbon, which is shown as white lines. Field lines of the $(x,z)$ components of the electric field have been superimposed to the plots. The plasmon energies are $\omega/\omega_0=3.19, 3.38, 4.30, 4.00, 4.01, 4.65$ for modes c-h, respectively, where $\omega_0=(e/\hbar)\sqrt{\langle|E_F|\rangle/W}$ [see Eq.\ (\ref{w0})].} \label{FigSM6}
\end{center}
\end{figure*}

As an example of calculation, we show in Fig.\ \ref{FigSM6}(a),(b) plasmon dispersion diagrams for both an inhomogeneously doped ribbon under the same conditions as in Fig.\ \ref{FigSM2} and a homogeneously doped ribbon. These plots are the same as Fig.\ 3(c),(d) of the main paper. We also show the near fields calculated for the three lowest-energy modes in each case, with $k_\parallel W=10$. These are plasmons of monopole (c,f), dipole (d,g), and quadrupole (e,h) character. In the inhomogeneous ribbon, the modes are associated with large field enhancement near the center. In contrast, the uniform ribbon hosts modes with large field enhancement near the edges.


\begin{thebibliography}{26}
\bibitem[{ana()}]{analimit}
\bibinfo{note}{This asymptotic expression can be easily derived by considering
  a thin biased wire placed above a backgate, and it might be useful to achieve
  large doping levels without electrical breakdown.}
\end{thebibliography}

\end{document}